\documentclass{emulateapj}
\usepackage{apjfonts}

\usepackage{graphics,graphicx,rotating}
\usepackage{natbib}
\usepackage{xspace}
\usepackage{amssymb}
\usepackage{amsmath}
\usepackage{epstopdf}
\usepackage{epsfig}
\usepackage{subfigure}

\shorttitle{Testing Hydrostatic Equilibrium in Clusters of Galaxies}
\shortauthors{Molnar, et al.}

\newcommand{\simless} 
     {\ensuremath{\lower 3pt\hbox{$\rlap{\raise5pt\hbox{$\char'074$}}\mathchar"7218$}}}
\newcommand{\simgreat}
     {\ensuremath{\lower 3pt\hbox{$\rlap{\raise5pt\hbox{$\char'076$}}\mathchar"7218$}}}

\newcommand{\simgt}{\lower.5ex\hbox{$\; \buildrel > \over \sim \;$}}
\newcommand{\simlt}{\lower.5ex\hbox{$\; \buildrel < \over \sim \;$}}

\newcommand{\nop}{{\noindent}}

\newcommand{\SUBARU}{{\sc Subaru}}
\newcommand{\CHANDRA}{{\sc Chandra}}
\newcommand{\SUZAKU}{{\sc Suzaku}}

\newcommand{\ENZO}{{\sc ENZO}}

\newcommand{\HE}{hydrostatic equilibrium}

\newcommand{\Pth}{{$P_{th}$}}
\newcommand{\Pnth}{{$P_{nth}$}}

\newcommand{\Pdyn}{{$P_{dyn}$}}
\newcommand{\Phe}{{$P_{he}$}}

\newcommand{\HMSUN}{{$h^{-1}\,\rm M_\odot$}}

\newcommand{\rmsub}[1]{\ensuremath{_{\rm #1}}}
\newcommand{\MSUN}{{\ensuremath{\mbox{\rm M}_{\odot}}}}
\newcommand{\RVIR}{\ensuremath{R\rmsub{vir}}}

\begin{document}

\title{Testing Strict Hydrostatic Equilibrium in Simulated Clusters of Galaxies: \\
                    Implications to Abell 1689}

\author{S. M. Molnar\altaffilmark{1}, I.-N. Chiu\altaffilmark{2,3}, 
             K. Umetsu\altaffilmark{1}, P. Chen\altaffilmark{2,3}, 
             N. Hearn\altaffilmark{4},  T. Broadhurst\altaffilmark{5,6,7}, 
             G. Bryan\altaffilmark{8} and C. Shang\altaffilmark{8}
}

\altaffiltext{1}{Institute of Astronomy and Astrophysics, Academia Sinica, 
                      P.O. Box 23-141, Taipei 106, Taiwan, R.O.C.; sandor@asiaa.sinica.edu.tw}

\altaffiltext{2}{Department of Physics, Institute of Astrophysics, \& Center
                      for Theoretical Sciences, National Taiwan University, Taipei 10617, Taiwan, R.O.C.}

\altaffiltext{3}{Leung Center for Cosmology and Particle Astrophysics, National Taiwan
                      University, Taipei 10617, Taiwan, R.O.C.}
                      
\altaffiltext{4}{Computational \& Information Systems Laboratory, 
                      National Center for Atmospheric Research, PO Box 3000, Boulder CO, 80305}

\altaffiltext{5}{School of Physics and Astronomy, Tel Aviv University, Israel}

\altaffiltext{6}
                {Department of Theoretical Physics, University of Basque Country UPV/EHU, Leioa, Spain}

\altaffiltext{7}
                {IKERBASQUE, Basque Foundation for Science,48011, Bilbao, Spain}

\altaffiltext{8}{Department of Astronomy, Columbia University, 550 West 120th Street, 
                      New York, NY 10027}

\begin{abstract}
Accurate mass determination of clusters of galaxies is crucial if they are to be used as 
cosmological probes.
However, there are some discrepancies between cluster masses determined
based on gravitational lensing, and X-ray observations assuming strict \HE\
(i.e., the equilibrium gas pressure is provided entirely by thermal pressure).
Cosmological simulations suggest that turbulent gas motions remaining from 
hierarchical structure formation may provide a significant contribution to the equilibrium
pressure in clusters.
We analyze a sample of massive clusters of galaxies drawn from high resolution 
cosmological simulations, and find a significant contribution (20\%--45\%) from non-thermal 
pressure near the center of relaxed clusters, and, in accord with previous studies, 
a minimum contribution at about 0.1 \RVIR, growing to about 30\%--45\% at the 
virial radius, \RVIR.
Our results strongly suggest that relaxed clusters should have significant non-thermal 
support in their core region.
As an example, we test the validity of strict \HE\ in the well-studied massive galaxy cluster 
Abell 1689 using the latest high resolution gravitational lensing and X-ray observations. 
We find a contribution of about 40\% from non-thermal pressure
within the core region of A1689, suggesting an alternate explanation for the 
mass discrepancy: the strict \HE\ is not valid in this region.
\end{abstract}


\section{Introduction}
\label{S:Intro}

Clusters of galaxies, the most massive virialized systems, form from the largest positive 
density fluctuations. The evolution of the abundance of these rare fluctuations are sensitive
to the cosmological model.
Also, the distribution of dark matter and gas in these systems provide a powerful 
test for our structure formation theories.
Mass determinations based on X-ray observations customarily assume spherical symmetry 
and strict \HE, i.e., the gas pressure is provided entirely by thermal pressure, 
$P_{he} = P_{th}$ (e.g., \citealt{Sara88}).
However, cosmological simulations suggest that even after equilibrium is established, 
a significant fraction of the pressure support against gravity comes from subsonic 
random gas motion in clusters
(Lau, Kravtsov \& Nagai 2009; \citealt{Maieet09,FanHumBuo09,YounBria07}; 
and references therein).

Previous studies focused on how cluster mass determinations are influenced
by non-thermal pressure support \citep{Zhanet10,Meneet09,LauKraNag09,Lagaet10}.
In this Letter, instead of determining the cluster mass, we focus on the 
dynamically important physical parameter of the intra-cluster gas (ICG): the pressure.
We use a sample of massive clusters of galaxies drawn from high resolution 
cosmological simulations and quantify the contribution from non-thermal 
pressure in relaxed clusters.

As an example, we test the validity of the common assumption of strict \HE\ in Abell 1689.
A1689 is one of the most thoroughly investigated 
massive galaxy clusters with a total mass of $1.5 \times 10^{15}$ \HMSUN, 
located at a redshift of 0.183 (\citealt{Coeet10,Penget09,Riemet09,Lemzet08,UmeBro08}; 
and references therein).
It has been found, under the assumption of spherical symmetry and 
strict hydrostatic equilibrium, that in the central region of A1689 the mass derived 
from X-ray observations is significantly lower than that inferred from gravitational 
lensing measurements (\citealt{Riemet09,Penget09,AndMad04}; and references therein).
We quantify the contribution from non-thermal pressure in A1689 
using the latest \CHANDRA, \SUZAKU\ and gravitational lensing observations, 
and make use of our results for simulated clusters to interpret the observations.
In the rest of the paper we assume a concordance cold dark matter (CDM) model 
with $\Omega _{m} = 0.3$, $\Omega _{\Lambda} = 0.7$, and $h=0.7$, 
where $h$ is defined as $H_0 = 100$ $h$ km s$^{-1}$Mpc$^{-1}$.
Errors, error bars and dashed lines represent 
$1\,\sigma$ confidence levels unless otherwise stated.

%
%
\begin{figure}
\centerline{
\epsfig{file=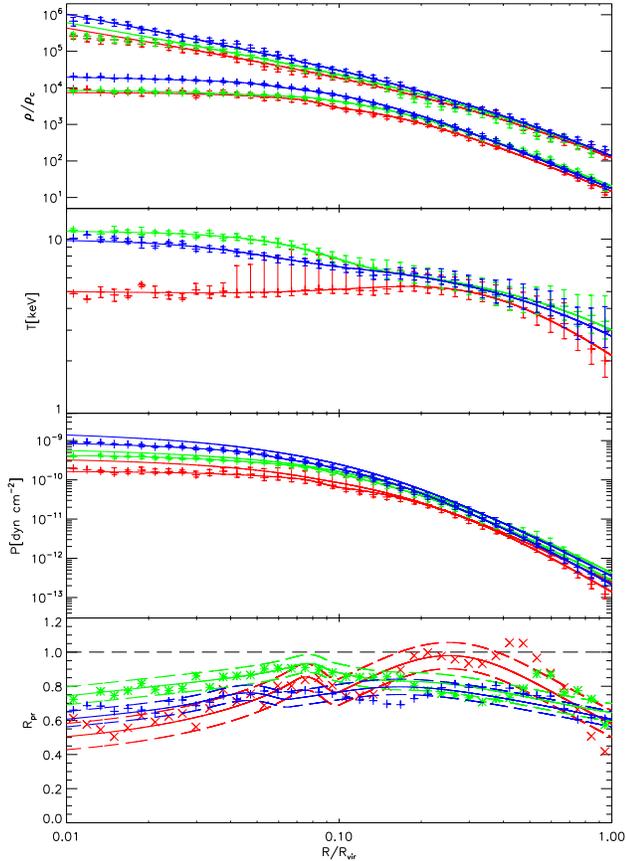,scale=0.5}
}
\caption{
 From top to bottom:
 radial profiles of dark matter and gas density (upper and lower curves) in 
 units of critical density, $\rho_c$; gas temperature (in keV);
 hydrostatic and thermal gas pressure (upper and lower curves, in dyn $\rm cm^{-2}$); 
 and pressure ratios, $P_{th}/P_{he}$, for massive simulated clusters.
Solid lines and points with error bars represent best fit models and data points 
for our relaxed clusters (blue: CL1; green: CL2), and for a cluster with a
non-relaxed core (red: CL3).
 We also show pressure ratios using gas pressure data points
 (blue plus signs, green stars and red crosses; see text for details).
\label{F:f1}
}
\end{figure} 

\smallskip
\section{Pressure in Simulated Clusters}
\label{S:HESIMU}

We derive the hydrostatic and thermal gas pressure profiles, $P_{th}$ and $P_{he}$, for massive relaxed
clusters drawn from cosmological adaptive mesh refinement (AMR) simulations 
performed with the cosmological code \ENZO\ \citep{OSheet04} assuming a spatially 
flat cosmological model very similar to the concordance model.
The AMR simulations were adiabatic in the sense that no heating, cooling, 
or feedback were included (for details see \citealt{YounBria07}).
The highest resolution was about 25 kpc at $r$ = 0.01 \RVIR\ 
(where the virial radius, \RVIR, is defined as in \citealt{BryaNorm98}), 
We selected relaxed clusters from the ten most massive clusters 
(1--2$\, \times 10^{15}~\MSUN$) based on having
a smooth spherically averaged density profile and no sign of recent major merger events 
(for details see \citealt{Molnet09}).

We show radial profiles for the two relaxed clusters (CL1 and CL2) 
and one cluster with a non-relaxed core (but otherwise relaxed, CL3) 
in Figure~\ref{F:f1}.
In this figure, the error bars represent the ${\it rms}$ of the density and temperature variations
due to angular averaging.
In the density and temperature plots 
the solid lines represent best fit models to the AMR data points assuming
double $\beta$ models for the electron density distribution, 
$n_e = n_1/(1+r^2/r_1^2)^{1.5 \beta_1} + n_2/(1+r^2/r_2^2)^{1.5 \beta_2}$, and 
$T_e = T_0  [ 1 + T_1 \exp \{ - (r/b_1)^{a_1} \} ] / (1 + r/b_2 )^{a_2}$
for the temperature.
We fixed $a_2 = 1.6$, which is an average value suggested by simulations \citep{YounBria07}.
We used $\rho_D \propto 1 / [ (r/d_1)^{1 + \alpha} (1 + r/d_1)^{3-\alpha} ]$
for the dark matter density profiles.

We derive the hydrostatic gas pressure, \Phe, using the equation of \HE\ 
assuming spherical symmetry, 

\begin{equation}  \label{E:HE}
      {d P_{he} (r) \over d r} = - { \rho_g(r) \, G\, M(r) \,  \over r^2}
,
\end{equation}
\nop
where $\rho_g(r)$ and $M(r)$ are the gas density and the cumulative total mass within the
3-dimensional (3D) radius, $r$ and $G$ is the gravitational constant.
We numerically integrate Equation~\ref{E:HE} using our fits.
The accretion shock maintains a finite value for $P_{he}$ at \RVIR, which we derive iteratively:
we integrate Equation~\ref{E:HE} inward from \RVIR\ demanding that the falloff of the 
pressure at large radii is a smooth function of the radius.
Note that, since the pressure drops more than three orders of magnitude from the cluster 
center to \RVIR, the pressure in the central region is insensitive to the choice of $P_{he}$
at \RVIR\ (within reasonable limits).

The resulting hydrostatic and thermal gas pressure profiles are shown in Figure~\ref{F:f1}.
In the 3rd panel the points show $P_{th}$. The solid lines connecting the points are 
not fits but derived from the density and temperature fits (and serve as a consistency test).
The sold lines above these lines represent the \HE\ pressure.
Note that, in the core region of CL3, the positive red error bars for the temperature are much 
larger than the negative ones. This is an indication of large positive deviations from the median 
in these radial bins, and a consequence of a non-relaxed core in CL3.
The pressure ratio, $R_{pr} = P_{th} / P_{he}$, profiles are also shown (bottom panel). 
In this panel, the errors (blue, green and red dashed lines) 
represent the average deviations of pressure ratio points
(plus signs, crosses and stars) from the mean (solid lines). 
We use these errors since they are larger than the statistical errors associated with our smooth 
fitting functions (except at the center of the cluster with the non-relaxed core, CL3).

\smallskip
\section{Testing Hydrostatic Equilibrium in Simulated Clusters}
\label{S:HEAMR}

We test the assumption of strict \HE\ in our simulated clusters using the pressure ratio 
profiles: $R_{pr} = P_{th} / P_{he}$.
These profiles in our relaxed simulated clusters show a similar trend
(Figure~\ref{F:f1}; bottom panel, blue and green lines).
In each cluster, $R_{pr}$ is small near the center ($r \simlt 0.05\,R_{vir}$), 
close to unity in $0.05 \,R_{vir} \simgt r \simlt 0.2 \,R_{vir}$, 
and decreases in the outer regions ($r \simgt 0.2 \, R_{vir}$).
Even in our cluster with a non-relaxed core, CL3, the average $R_{pr}$ shows a 
similar trend (red lines).

Quantitatively, we find that the contribution from non-thermal pressure, $P_{nth} = P_{he} - P_{th}$, 
is decreasing from 20\%--45\% near the cluster center (0.01\RVIR), reaches a minimum 
of 5\%--30\% at about 0.1\RVIR, and increases up to about 30\%--45\% at \RVIR. 
In the region of overlap with other simulations, $r \simgreat 0.1 \, R_{vir}$, 
similar contributions have been found in clusters from \Pnth: 5\%--15\% 
at about  0.1\RVIR, increasing with radius to 20\%--40\% at \RVIR\ \citep{LauKraNag09}.
Our results for non-thermal pressure support are also consistent with those of \cite{YounBria07}.
Thus we find that in the central regions of our massive clusters, \Pth\ is significantly less 
than \Phe\ (Figure~\ref{F:f1}).

In our adiabatic simulations, this non-thermal pressure support is provided by subsonic random 
gas motion, i.e., $P_{dyn} = P_{nth}$, where \Pdyn\ 
is the dynamical pressure (e.g., \citealt{LauKraNag09}). 
We show the average dynamical pressure ratio, $\langle P_{dyn}/P_{he} \rangle$,
profile in Figure~\ref{F:f3} (black lines).
We estimate the errors in $\langle P_{dyn}/P_{he} \rangle$ as $Min\{ (R_{pr} - 1\sigma)_i \}$
and $Max\{ (R_{pr} + 1\sigma)_i \}$ (black dashed lines). 
We conclude that in the central regions of massive simulated clusters there is a significant
contribution from dynamical pressure.

Based on our results, we find that there is no need for feedback from a central AGN for 
the strict \HE\ to break down in the central regions of clusters.
This break down is due to subsonic random gas motion, which 
is a direct consequence of hierarchical structure formation.
Near \RVIR, the contribution from $P_{nth}$ may reach 30\%--45\%. 
This is probably due to more recent and still ongoing slow accretion since the outer regions 
of clusters have not reached equilibrium due to the large sound crossing time, 
about 1 Gyr (for a more detailed analysis see \citealt{Kawaet09}).
Our results for simulated clusters strongly suggest that even relaxed clusters should 
have significant non-thermal support in their core regions.

%
%
\begin{figure}[placement t]
\centerline{
\epsfig{file=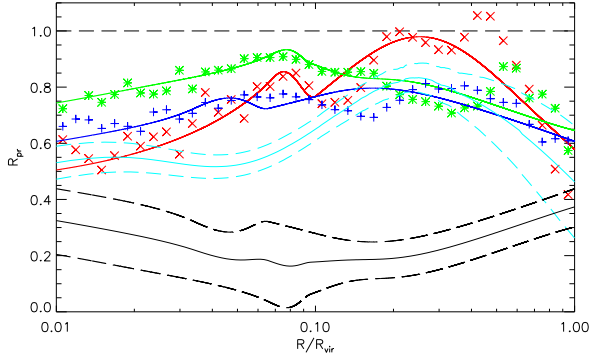,scale=0.5}
}
\caption{
 Pressure ratio profiles, $P_{th}/P_{he}$ derived for simulated massive clusters
 CL1, CL2 and CL3 (code is the same for the colored solid liens and points 
 (as in Figure~\ref{F:f1}), 
 and for A1689 using the latest NFW lensing model (cyan lines, same as in 
 Figure~\ref{F:f3}).
 We also show the average dynamical pressure ratios, $\langle P_{dyn} /P_{he} \rangle$,
 for simulated relaxed clusters CL1 and CL2 with errors (black lines). 
 See text for details and the definition of errors (dashed lines).
\label{F:f2}
}
\end{figure} 

%
%
\begin{figure}
\centerline{
\epsfig{file=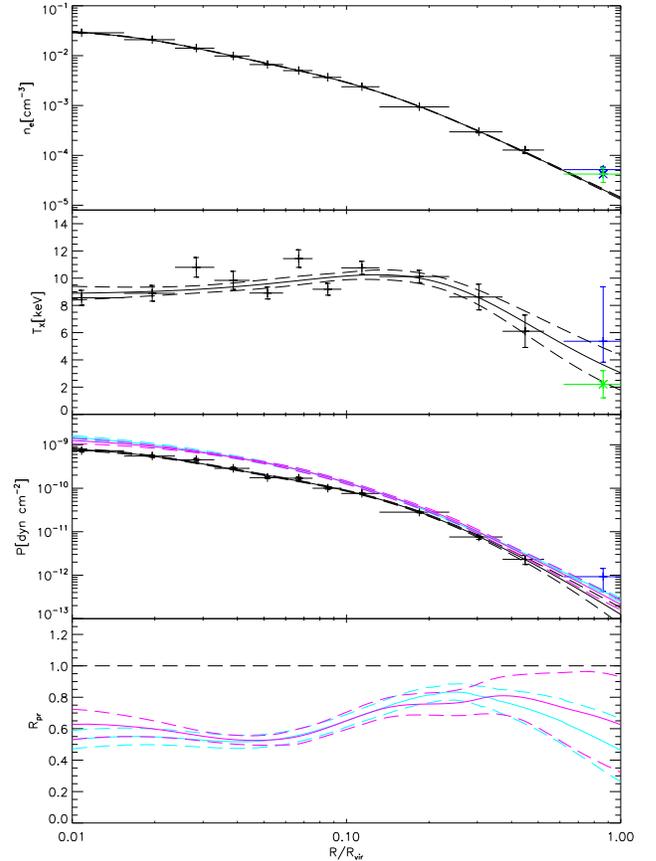,scale=0.5}
}
\caption{
 Top 3 panels: radial profiles of electron density and temperature, 
 and thermal pressure in A1689 
 derived from \CHANDRA\ observations (black points with error bars), 
 and adopted from \SUZAKU\ observations 
 (blue points with error bars; offset1; Kawaharada et al. 2010).
 On the op 2 panels we also show the best fit models (black solid lines),
 and, as a reference, the last point of offset 2 of \SUZAKU\ 
 (green stars with error bars). 
 In the 3rd panel, the black solid line is
 derived from the density and temperature fits. 
 In this panel, we also show hydrostatic equilibrium pressure, \Phe, 
 derived from gravitational lensing 
 using an NFW and a non-parametric model for the total mass distribution 
 (cyan and magenta solid lines). 
 Pressure ratios, $R_{pr} = P_{th}/P_{he}$, are shown in the
 bottom panel (color code is the same as in the 3rd panel; 
 horizontal black dashed line represents ratio of unity).
\label{F:f3}
}
\end{figure} 

\smallskip
\section{Testing Hydrostatic Equilibrium in A1689}
\label{S:A1689HE}

As an example, we determine the pressure profiles in the well studied massive galaxy cluster A1689. 
We derive $P_{th}$ from our analysis of the three longest
publicly available \CHANDRA\ ACIS-I observations of A1689 (ObsIDs: 5004, 6930 and 7289).
After following standard ACIS data 
preparation\footnote{$http://cxc.harvard.edu/ciao/guides/acis\_data.html$}, 
the total exposure time of 171 ksec was reduced to 128.7 ksec. 
We fixed the redshift and the photoelectric absorption at the galactic value, 
and used $n = n_e + n_H $, where $n_e$ and $n_H$ are the electron and 
hydrogen densities assuming $n_e / n_H = 1.1737$ (as in \citealt{Penget09}).
Since these observations were taken in the Very FAINT telemetry mode, 
while the FAINT mode was used for the background observations,
we chose to use local background.
Assuming spherical symmetry, we determined the de-projected 3D densities 
and electron temperatures, $T_e$, out to 0.6 \RVIR\ for each shell fitting all shells 
and data sets simultaneously using XSPEC-12.5 applying C-statistic
(Figure~\ref{F:f3}; black points with error bars). 
Our results are consistent with those Peng et al. (2009; see their Figure 11).
We calculate the thermal pressure in A1689 using the ideal gas law.
For $r > 0.6$ \RVIR\ we show the results for offset 1 from the \SUZAKU\ observations 
\citep{Kawaet09}, which we use in our analysis.
We also show the results for offset 2 (green points)
to represent the similar results from offsets 2, 3 and  4 of \SUZAKU.
The lower resolution \SUZAKU\ results agree with those from \CHANDRA\ 
for $r < 0.6$ \RVIR, thus we use only the last radial data points from \SUZAKU.
We use offset 1 because the other offsets would give us even higher 
non-thermal pressure contribution due to the very low temperatures for 
offsets 2, 3, and 4 relative to offset 1.

We derive the hydrostatic gas pressure, \Phe, numerically integrating Equation~\ref{E:HE} 
assuming the same gas density and temperature models as in Section~\S\ref{S:HEAMR}.
We find the best fit parameters for the density:
$n_1 = 3.0\times10^{-2}\pm1.1\times10^{-3}\,\rm cm^{-3}$, 
$n_2 = 6.0\times10^{-3}\pm2.7\times10^{-4}\,\rm cm^{-3}$, 
($r_1,\beta_1$) = ($7.5\times10^{-2}\pm3.8\times10^{-3},1.01\pm6.0\times10^{-2}$),                        
($r_2,\beta_2$) = ($3.1\times10^{-1}\pm8.8\times10^{-3},0.90\pm0.18\times10^{-3}$);
and for the temperature: $T_0 = 17.1\pm3.6$ keV, $T_1 = -0.48\pm0.12$ keV, 
$(b_1,a_1)$ = ($0.40\pm0.15,1.3\pm0.73$), and $(b_2,a_2)$ = ($2.8\pm2.3,2.3\pm2.2$), 
where all scale radii, $r_1,r_2,b_1,b_2$, are in Mpc (Figure~\ref{F:f3}, black solid lines).
The fitted parameters determining the large scale temperature profile, $(b_2,a_2)$,
have large errors due to the large uncertainty in the \SUZAKU\ 
temperature measurement near \RVIR.
We derive the gas density using $\rho_g = n_e \, \mu_e \, m_P$, where 
$m_P$ is the proton mass (we adopt $\mu_e = 1.146$; see e.g., \citealt{Penget09}).
We show the thermal gas pressure, \Pth, derived from the best fit density and temperature 
profiles in Figure~\ref{F:f3} (3rd panel; black line).
As a consistency check, we also derive the thermal pressure directly from the 
deprojected density and temperature points (points with error bars, same panel).
We determine the cumulative mass function, $M(r)$
using the best-fit spherical Navarro, Frenk \& White (NFW) model \citep{NFW97}
based on the latest high resolution mass modeling of A1689 by \cite{Coeet10}. 
As a comparison, we also use $M(r)$ based on a non-parametric deprojection 
of the 2D mass distribution derived from a joint strong and weak lensing analysis of 
{\it Hubble Space Telescope} Advanced Camera for Surveys (ACS) and \SUBARU\ 
observations of A1689 \citep{UmeBro08} 
assuming spherical symmetry (see also \citealt{Kawaet09}).
Our results for $P_{he}$ are shown in Figure~\ref{F:f3} (panel 3).
The errors for the NFW model were estimated using Monte Carlo simulations assuming a 
Gaussian probability distribution for the parameters centered on their respective 
best fit values. 
The errors for the non-parametric model were estimated from Monte Carlo simulations 
based on the full covariance matrix of the lensing convergence profile 
(for details see \citealt{Umetet10}).

The pressure ratios derived from observations of A1689,
are shown in Figure~\ref{F:f3} (bottom panel).
These ratios, based on NFW and a non-parametric model for $M(r)$, 
are consistent with each other (Figure~\ref{F:f3}; cyan and magenta lines).
Similarly to our simulated clusters, we find a significant 
contribution from \Pnth\ in the core region and near \RVIR\ (Figure~\ref{F:f3}).
The contribution from \Pnth\ is about 40\%$\pm$10\% in A1689 out to about 0.1\RVIR,
which is somewhat smaller than that in simulated clusters, 
although, at the center, it is within the errors due to cluster-to-cluster scatter.
Also, from this figure we see that the maximum of the average ratio in
simulated clusters is located somewhat closer to the center than the maximum 
in A1689 (Figure~\ref{F:f3}).

However, we expect that in the core region of A1689 non-gravitational processes 
are also important due to feedback from active galactic nuclei (\citealt{Mccaet09}; 
and references therein). 
In general, contributions to the pressure in clusters may come from turbulence, 
magnetic fields and cosmic rays (e.g., \citealt{Vazzet09AA504}).
An attempt to separate these non-thermal pressure components based on their 
assumed functional form derived from numerical simulations has been carried out 
by Lagana et al. (2009). However, as they pointed out, this decomposition 
strongly depends on the assumed functional forms of the components.
Unfortunately, at this point, we cannot determine the dominant contribution
to the non-thermal pressure support in A1689 from observations directly.

\smallskip
\section{Conclusion}
\label{S:Conclusion}

We have analyzed massive clusters of galaxies drawn from high resolution 
cosmological simulations and quantified the non-thermal pressure support in 
relaxed clusters with high resolution from $r \approx 0.01$\RVIR\ out to \RVIR.
We have found a significant contribution from non-thermal pressure in
simulated clusters due to subsonic random gas motion:
20\%--45\% at $r \approx 0.01$\RVIR\ and 30\%--45\% at \RVIR\ having
a minimum support of 5\%--30\% at $r \approx 0.1$\RVIR\ (Figure~\ref{F:f1}).
Our results strongly suggest that relaxed clusters should have significant non-thermal 
support in their core region, and that this non-thermal
pressure support should be taken into account when analyzing clusters.

As a test case, we have quantified the non-thermal pressure
support in the well studied galaxy cluster A1689.
Based on our results for the thermal gas pressure and the hydrostatic 
equilibrium pressure determined from X-ray and gravitational lensing 
observations of A1689, assuming spherical symmetry, we have found a significant, 
40$\pm10$\%, contribution from non-thermal pressure within 0.1 \RVIR.
We conclude that the mass discrepancy in the central region of A1689 
can be explained if we assume that the strict \HE\ is not valid in this region.
We need to test the assumption of \HE\ in more clusters to find out how common 
this large amount of non-thermal pressure contribution in their core region.

While our spherical models for A1689 take into account support from non-thermal gas 
pressure as suggested by CDM models, they predict a high concentration parameter, 
the triaxial models of \cite{MorPedLim10}, \cite{CorKinClo09} and \cite{Sereet06ApJ645} 
do not take into account non-thermal pressure but provide a concentration parameter 
consistent with the predictions of CDM models. 
\cite{Penget09} have found that a prolate gas distribution could solve the mass 
discrepancy, but it overestimates the total mass at large radii significantly, and implies 
a larger ellipticity than predicted by CDM models.
Although all of these models solve the mass discrepancy in A1689,
neither of them is fully consistent with all predictions of CDM models.
Our results suggest that a physical cluster model for A1689 
with a triaxial mass distribution including support from non-thermal pressure 
might be fully consistent with all observations and the predictions of CDM models.

\acknowledgements
We thank the anonymous referee for comments and suggestions which helped to 
improve on the presentation of our results substantially, M. Birkinshaw for enlightening 
discussions, and N. Okabe for discussions of the \SUZAKU\ observations of A1689.
This work was supported in part by National Science Foundation Grants No. AST-05-07161 
and AST-05-47823, and supercomputing resources from the National Center for 
Supercomputing Applications.

%
%
\bibliographystyle{apj}

 
\end{document}